\documentclass[12pt]{article}
\pdfoutput=1

\usepackage[T1]{fontenc}
\usepackage{amsmath,amssymb,graphicx} 
\usepackage{epsf}
\usepackage{tikz}
\usepackage{pstricks}
\usepackage{feynmp}
\usepackage{cite}
\usepackage{multirow}

\newcommand{\beq}{\begin{eqnarray*}}
\newcommand{\eeq}{\end{eqnarray*}}
\definecolor{orange}{rgb}{1, 0.4, 0} 
\definecolor{vertfonce}{rgb}{0, 0.4, 0} 
\definecolor{marron}{rgb}{0.36,0.13,0.00} 
\definecolor{purple}{rgb}{0.4,0.0,0.4} 
\definecolor{pink}{rgb}{0.8,0.3,0.6} 
\definecolor{gray}{rgb}{0.3,0.3,0.3}

\begin{document}
\begin{fmffile}{fgraphs}

\begin{titlepage}
\begin{flushright}

Preprint no. LYCEN-2015-04
\end{flushright}

\vskip.5cm
\begin{center}
{\huge \bf 
Higgs couplings and BSM physics: Run I Legacy constraints}

\vskip.1cm
\end{center}
\vskip1cm
\begin{center}
{\bf {Jean-Baptiste Flament}}
\end{center}
\vskip 8pt

\begin{center}
{\it
Universit\'e de Lyon, F-69622 Lyon, France; Universit\'e Lyon 1, Villeurbanne;\\
CNRS/IN2P3, UMR5822, Institut de Physique Nucl\'eaire de Lyon\\
F-69622 Villeurbanne Cedex, France} \\
\vspace*{.5cm}
{\tt  j-b.flament@ipnl.in2p3.fr}
\end{center}

\vskip2truecm

\begin{abstract}
\vskip 3pt
\noindent

We consider the Higgs boson decay processes and its production including all Run I results, through a parametrisation tailored for testing 
models of new physics beyond the Standard Model, and complementary to the one used by the LHC working groups. Different formalisms 
allow to best address different aspects of the Higgs boson physics. The choice of a particular parametrisation depends on a non-obvious 
balance of quantity and quality of the available experimental data, envisaged purpose for the parametrisation and degree of model independence, 
importance of the radiative corrections, and scale at which new particles appear explicitly in the physical spectrum. 

The most refined constraints can only be obtained by the experimental collaborations at present, as information about correlation between the various uncertainties on the different decay modes is not completely available in the public domain. It is therefore important that different approaches are considered and that the most detailed information is made available to allow testing the different aspects of the Higgs boson physics and the possible hints of physics beyond the Standard Model.

\end{abstract}

\end{titlepage}

\section*{Introduction}

After Run I and with the break in data acquisition for the 2013-2014 upgrade to nominal energy, LHC physics experimentalists have further finalized the analysis of data that was gathered throughout Run I (2011-2012), which led in their preliminary steps to the announcement of the discovery of the 125 GeV scalar boson in July of 2012. As of April of 2015, the CMS and ATLAS collaboration have released most final results concerning the study at 7 and 8 TeV of the couplings of the detected boson to the rest of the Standard Model (SM) particles, and are ready to start acquiring data at a higher centre-of-mass energy. 

In the Standard Model, the Higgs boson, physical particle of the Higgs doublet after the breaking of the electroweak symmetry, is of course coupled to the weak gauge bosons and fermions to which the mechanism effectively gives mass at low energies. At loop level, two final states experimentally stand out of the many possible Higgs decays: the one with two photons and the one with two gluons.
Those are therefore the main interesting channels that are looked for in colliders, and the couplings involved can be used to constrain new physics models in which they are modified. 
Models beyond the Standard Model (BSM) describing the electroweak symmetry breaking often do so in a different way than the SM, implying a change in many of these couplings with respect to the SM predictions. Instead of doing computations and extracting constraints individually for each model, it is convenient to use model-independent parametrisations of the couplings, in order to perform the whole constraining procedure a few steps shorter. In \cite{cacciapaglia_0901}, the authors advocated for the use of a parametrisation tailored for the study of models where loops have a sizeable impact on effective Higgs couplings. 

\section{Higgs rescaled couplings : a loop oriented parametrisation}

Among the particle to which the Higgs boson couples, a distinction can be made between particles coupling at tree level and particles couplings only through loops of other particles.

\subsection{Tree-level couplings}

\begin{center}
   \fmfframe(0,14)(0,14){
    \begin{fmfgraph*}(60,40)
    \fmfleft{p1}
    \fmfright{p2,p3}
    \fmflabel{$H$}{p1}
    \fmflabel{$q,l$}{p2}
    \fmf{dashes}{p1,v1}
    \fmf{fermion}{v1,p2}
    \fmf{fermion}{p3,v1}
    \fmffixed{(25,0)}{p1,v1}
    \end{fmfgraph*}\hspace{2cm}
    \begin{fmfgraph*}(60,40)
    \fmfleft{p1}
    \fmfright{p2,p3}
    \fmflabel{$H$}{p1}
    \fmflabel{$V$}{p2}
    \fmf{dashes}{p1,v1}
    \fmf{boson}{v1,p2}
    \fmf{boson}{v1,p3}
    \fmffixed{(25,0)}{p1,v1}
    \end{fmfgraph*}
    }
\end{center}

In the fundamental SM Lagrangian the Higgs doublet is coupled to elementary fermions and gauge bosons, and these couplings translate into those of the Higgs boson after symmetry breaking. The latter can therefore be defined straightforwardly as the coefficient in front of the coupling terms in the SM effective Lagrangian.


In a BSM model however, whether the Higgs mechanism is still relevant and additional physics come into play, or the whole symmetry breaking occurs in a different way, the Lagrangian may be different (effectively or fundamentally). Those couplings are therefore modified, and we can parametrise this modification through the definition of the ratio:

\begin{displaymath}
 \kappa_{i} = \frac{g_{hii}}{g_{hii}^{SM}}
\end{displaymath}

Parameters defined in this way (in our case $\kappa_b,\ \kappa_t,\ \kappa_\tau, \text{ and } \kappa_V$, respectively rescaling the Higgs couplings to down-type quarks, up-type quarks, charged leptons, and weak vector bosons) obviously have a value of 1 in the SM, and may deviate from this value when the model is different from the SM.

\subsection{Loop-induced couplings}

\begin{center}
   \fmfframe(0,14)(0,14){
    \begin{fmfgraph*}(60,40)
    \fmfleft{p1}
    \fmfright{p2,p3}
    \fmflabel{$H$}{p1}
    \fmflabel{$\gamma,g$}{p2}
    \fmflabel{$\gamma,g$}{p3}
    \fmf{dashes}{p1,v1}
    \fmf{photon,width=1pt}{v2,p2}
    \fmf{photon}{v3,p3}
    \fmf{plain,tension=0.5}{v1,v2}
    \fmf{plain,tension=0.2}{v2,v3}
    \fmf{plain,tension=0.5,label=$t',,\tilde{t}...$}{v3,v1}
    \fmffixed{(15,0)}{p1,v1}
    \end{fmfgraph*}
    }
\end{center}

In order to use the experimental constraints from those channels to test a model, it is also convenient to use simple parameters allowing comparison between experimental results and model predictions. However, as stated previously, the Higgs couplings to pairs of photons or of gluons do not appear explicitly in the fundamental Lagrangian but only appear in an effective description when considering the effects of loops. Therefore, the parameters are trickier to define in a unique way. One of the standard approaches is to do the same as in the case of direct couplings and rescale the effective couplings of the model to the same effective couplings in the SM.

However, this approach does not distinguish the variation of this effective coupling due to the change of the direct couplings from the one due to additional loops coming from the new physics.
Another approach, introduced in \cite{cacciapaglia_0901} is interesting in the case of models where additional particles appear. In this case, one can take into account the loop-induced character of these couplings and introduce the change in the effective coupling at the level of the loop calculation.
More precisely, in the SM one can compute the partial width of the Higgs in those channels through:

\begin{eqnarray*}
 \Gamma_{\gamma\gamma} &=& \frac{G_F \alpha^2 m_H^3}{128 \sqrt{2} \pi^3} \left|  A_W (\tau_W) + 
C^\gamma_t \; 3 \left( \frac{2}{3} \right)^2 A_t (\tau_t) + \dots \right|^2\,, \\
\Gamma_{g g} &=& \frac{G_F \alpha_s^2 m_H^3}{16 \sqrt{2} \pi^3} \left| C^g_t \frac{1}{2} A_t (\tau_t) + \dots \right|^2
\end{eqnarray*}

where $A_W$ and $A_t$ represent the analytical amplitude of the $W$ boson and top quark loops, $C^{\gamma}_t$ and $C^g_t$ include QCD corrections, and the dots represent negligible lighter fermions contributions. Defining $\tau = \frac{m_H^2}{4 m^2}$, the relevant expressions are:

\beq
A_t (\tau) &=& \frac{2}{\tau^2} \left( \tau + (\tau-1) f (\tau) \right)\,, \\
A_W (\tau) &=& - \frac{1}{\tau^2} \left( 2 \tau^2 + 3 \tau + 3 (2 \tau - 1) f (\tau) \right)\,, \\
\text{and }f (\tau) &=& \left\{ 
\begin{array}{lc}
\mbox{arcsin}^2 \sqrt{\tau}  & \tau \leq 1 \\
- \frac{1}{4} \left[ \log \frac{1+\sqrt{1-\tau^{-1}}}{1-\sqrt{1-\tau^{-1}}} - i \pi \right]^2 & \tau > 1 
\end{array} \right.\,.
\eeq

Additional loop diagrams coming from new physics contribute to these amplitude, and we therefore define the parameters $\kappa_{\gamma\gamma}$ and $\kappa_{gg}$ as:

\begin{center}
\parbox{5 cm}{$$\kappa_{\gamma\gamma} = \frac{A^\gamma_{NP}}{C^\gamma_t \; 3 \left( \frac{2}{3} \right)^2 A_t\left(\tau_t\right)} $$}     \parbox{5cm}{$$\kappa_{gg} =\frac{A^g_{NP}}{C^g_t\, \frac{1}{2}\, A_t\left(\tau_t\right)} $$}
\end{center}

which gives for the partial widths in a new physics model:

\beq
\Gamma_{\gamma \gamma} &=& \frac{G_F \alpha^2 m_H^3}{128 \sqrt{2} \pi^3} \left|  \kappa_V \, A_W (\tau_W) + 
C^\gamma_t \; 3 \left( \frac{2}{3} \right)^2 A_t (\tau_t)\; [\kappa_t+\kappa_{\gamma \gamma} ] + \dots \right|^2\,, \\
\Gamma_{g g} &=& \frac{G_F \alpha_s^2 m_H^3}{16 \sqrt{2} \pi^3} \left| C^g_t \frac{1}{2} A_t (\tau_t)\; [\kappa_t+\kappa_{gg}] + \dots \right|^2\,,
\eeq

The first thing one can notice here is that the contributions to the variation of the width from the modification of the tree-level coupling and from the existence of new loops are here decorrelated, since the first are taken into account thanks to the $\kappa_V$ and $\kappa_t$ parameters. On the other side, the normalisation to the top amplitude is arbitrary, but stays general and might gives interesting values for models where the new physics include top partners.

As a side comment, we can note that in the case where the custodial character of the electroweak symmetry breaking is questioned (\textsl{i.e.} when couplings to the $W$ and $Z$ bosons are no longer rescaled by a common parameter), there is an advantage in considering the separate amplitude calculation rather than a simple width rescaling, as the parameter rescaling the amplitude of the W loop can then be described as $\kappa_W$. In this prospect, additional information can be obtained in this channel from the use of this parametrisation, as  $\kappa_W$ appears explicitly in the expression of $\Gamma_{\gamma \gamma}$, as opposed to $\kappa_Z$. However, as we will see in the second part, most LHC published results use a common rescaling factor for $W$ and $Z$ bosons couplings, making this information impossible to recast here, as it is only accessible to experimentalists.

On another hand, the parameter describing the loops coupling the Higgs boson to a pair of gluons can also be used to describe the modification of the cross section of production of Higgs through the fusion of gluons from the colliding protons, which is the main production mode at hadron colliders. We consider here that the effective coupling is the same for production cross section and decay width, although this might not be the case due to PDF effects.

\subsection{Interpretation of BSM models}

In particle physics experiments, measurements rely on the counting of events, therefore depending on differential cross sections, which can often be calculated in given models. As a consequence, the couplings involved in the calculation of cross sections have an influence on the expected result.

More specifically, one can define the \textsl{signal strength} $\mu$ as the number of events expected in a model for a given signal normalized to the number of events expected in the SM:  $n_{s}^{NP}= \mu\ n_s^{SM}$. They are therefore functions that depend on as many parameters as the cross sections. However, if the additional physics is expected not to change the shape of the cross-section distributions, it is reasonable to consider only one constant signal strength per measured physical process. Same as the cross sections, the signal strengths depend on the values of the couplings, and we can therefore write them as functions of our rescaling parameters:

\begin{displaymath}
 \mu\left(\kappa_V,\kappa_t,\kappa_b,\kappa_\tau,\kappa_{\gamma\gamma},\kappa_{gg}\right) = \frac{n^{NP}\left(\kappa_V,\kappa_t,\kappa_b,\kappa_\tau,\kappa_{\gamma\gamma},\kappa_{gg}\right)}{n^{SM}} =\frac{\sigma^{NP}}{\sigma^{SM}}
\end{displaymath}

\section{Recasting Higgs constraints}

\subsection{LHC data}

In the case of the Higgs boson analyses, the previous definitions would provide us with one signal strength per combination of production mode and measured decay channel. 

However, several assumptions make this number decrease. 
First, under the narrow width approximation (which experimentally seems valid), the cross sections for a given process can be factorized as the Higgs production cross section times the branching ratio of the final state. We can therefore use only one rescaling factor per production mode and one per decay channel.
Furthermore, as VBF and VH production modes rely at tree level on the Higgs coupling to weak gauge bosons, we can expect them to be rescaled in the same way. On the same level of idea, in the SM the $ggh$ production mode relies mostly on a top quark loop, and therefore scales in the same way as the $t\bar{t}h$ production mode. Although this is not exactly true (or we wouldn't have included a $\kappa_{gg}$ parameter), there is no analysis at the moment that include both the $ggh$ and $t\bar{t}h$ production modes (the $t\bar{t}h$ production mode is only probed through Higgs decays to $b\bar{b}$).

One of the forms under which the experimental collaborations provide the results of their analyses is as exclusion contours for given confidence levels, in the plane mapped by the production modes signal strengths $\mu_{VBF/VH}$ and $\mu_{ggh/t\bar{t}h}$. Each studied final state gives an exclusion contour, corresponding to the constraints on the signal strength of the whole process, from production to decay. 

Fig. 5 \textsl{in} \cite{Khachatryan:2014jba} summarises the final results of the different analyses for $7$ and $8\ TeV$ of the CMS collaboration, gathering the various exclusion contours at $68 \%\ C.L.$ for the various decay channels.

In the case of the ATLAS collaboration, a single figure gathering the contours obtained from each analysis after the legacy analysis of Run I was not published (as of writing of this paper). Therefore, the individual relevant figures were gathered from the analysis papers ($\gamma\gamma$, Fig. 20 \textsl{in} \cite{PhysRevD.90.112015}, $\tau^+\tau^-$ Fig. 12 \textsl{in} \cite{Aad:1982276}, $WW^\ast$ Fig. 40 \textsl{in} \cite{Aad:1975394} and $ZZ^\ast\to 4\ell$ Fig. 20 \textsl{in}\cite{PhysRevD.91.012006})

As an addition, ATLAS results from the $b\bar{b}$ Higgs decay searches were included as an exclusion band around a central value (Fig. 20 \textsl{in}\cite{Aad:bbATLAS}). This comes from the fact that this analysis relies on a selection tailored for events where the Higgs boson is produced through the VH production mode, using the presence of the additional gauge boson to reduce QCD noise. As a consequence, this analysis does not constrain $\mu_{ggH/t\bar{t}H}$.

\subsection{Fitting contours as ellipses}

As these exclusions take into account possible statistical fluctuations, experimental and theoretical systematic uncertainties on the measurements, one can assume that the likelihoods of the signal strengths, for large numbers of measured events, are distributed as Gaussians. This means that the log-likelihood functions are paraboloids. In terms of constraints, this can be translated into the fact that the exclusion contours at a given confidence level are ellipses.

\begin{figure}[ht] 
\begin{center}
\includegraphics[width=0.45\textwidth]{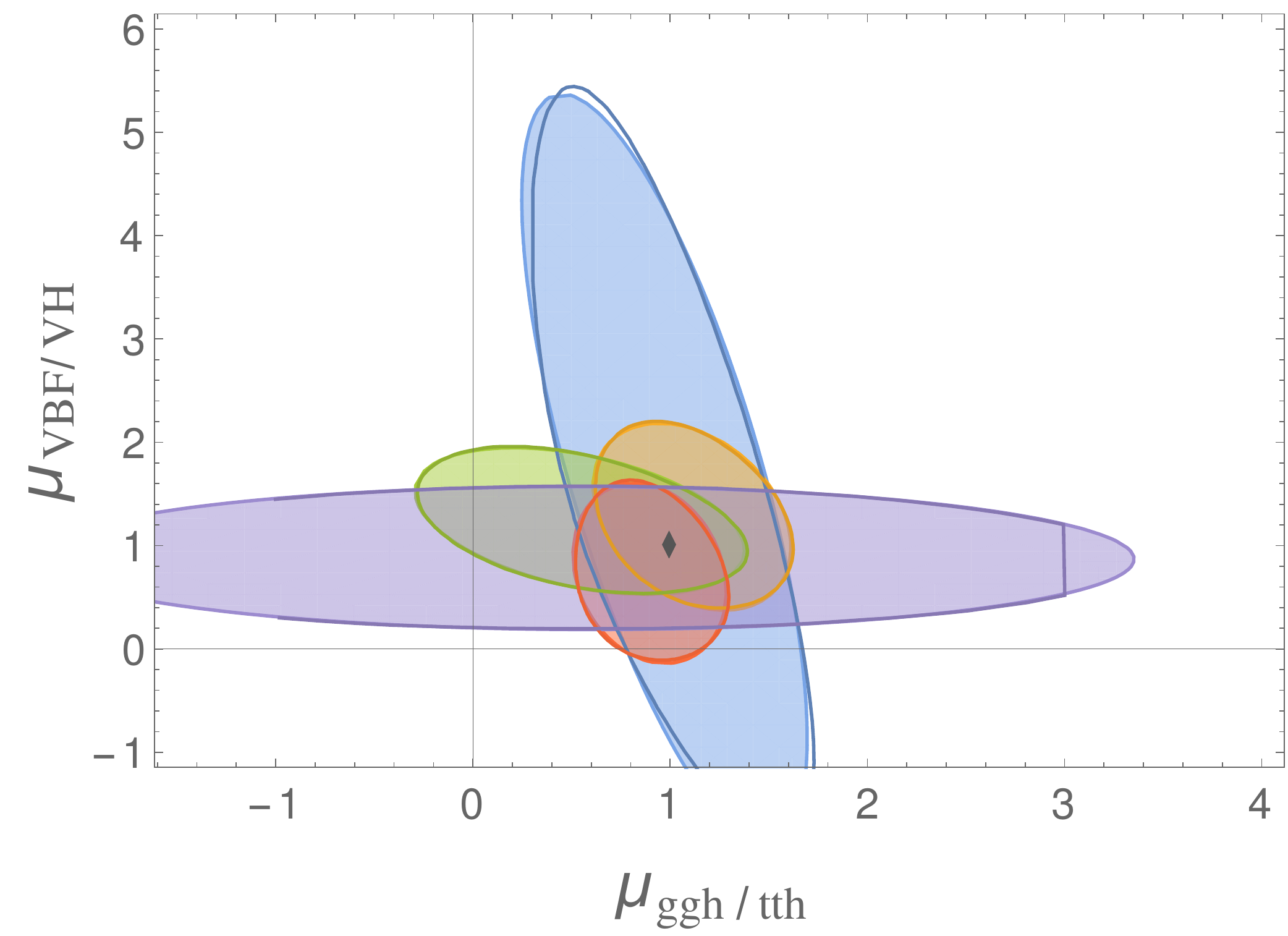} ~~~~~~
 \includegraphics[width=0.45\textwidth]{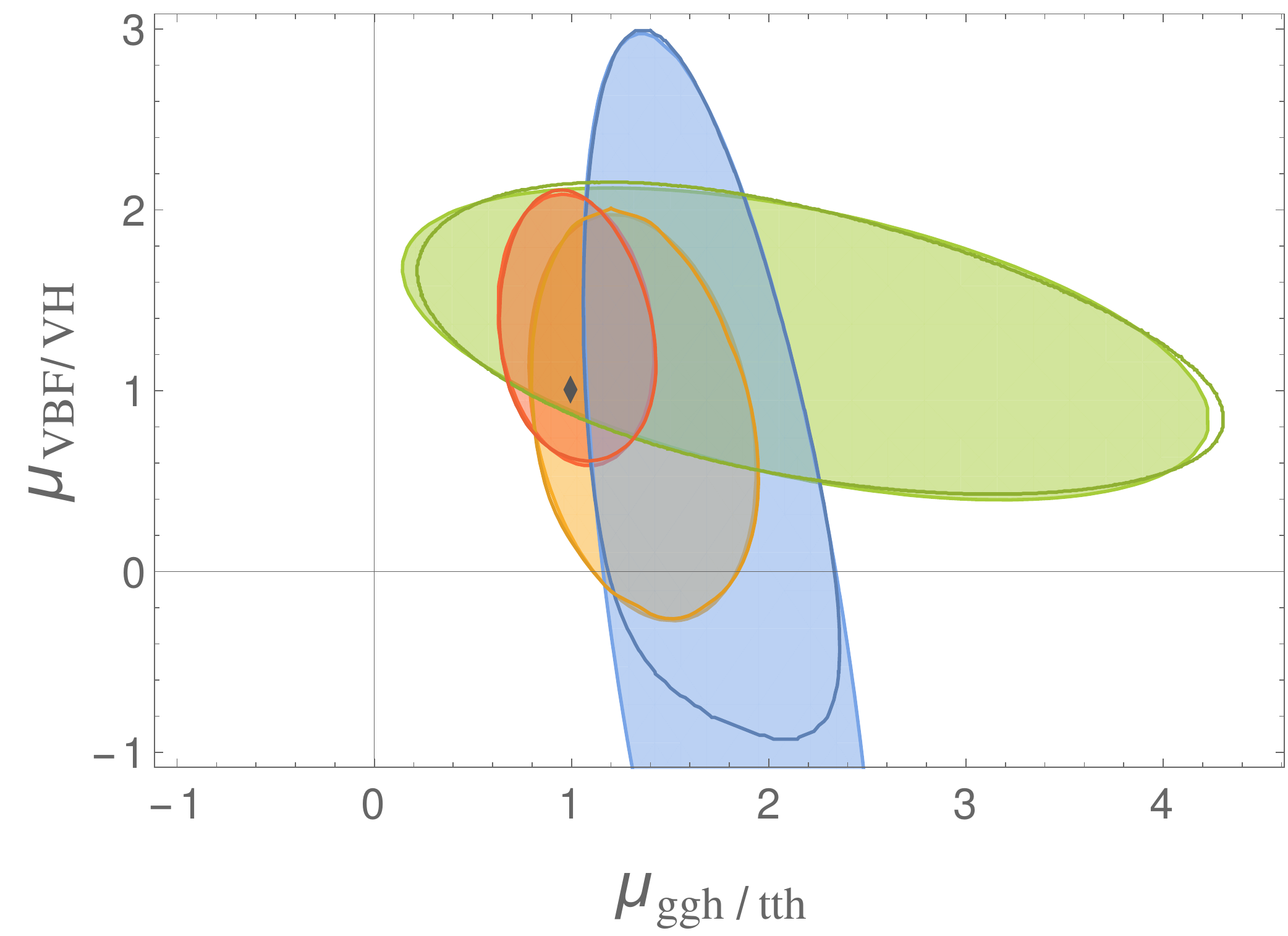}
\caption{Relevance of the hypothesis of a Gaussian likelihood: comparison of experimental $68 \% C.L.$ exclusion contours with the fitted contours (right: CMS data, left: ATLAS data). Colour code per final state: purple: $b\bar{b}$, yellow: $\gamma\gamma$, green: $\tau\tau$, red: WW, blue: ZZ. Gray rhombus: SM}
\label{fig:Fit1sig}
\end{center}
\end{figure}

\begin{figure}[ht]
\begin{center}
\includegraphics[width=0.45\textwidth]{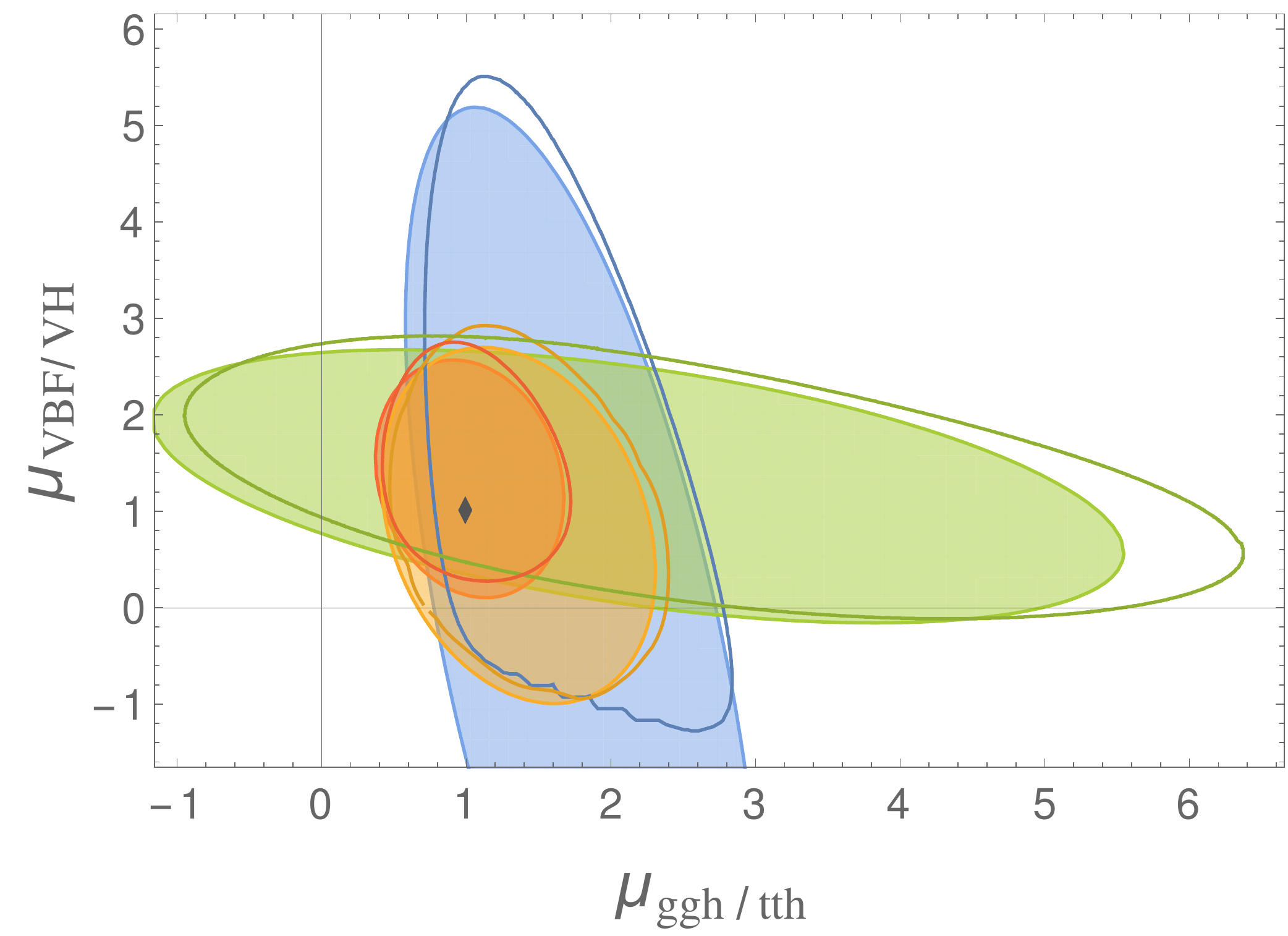}
\caption{Relevance of the  hypothesis of a Gaussian likelihood: comparison of experimental $95 \% C.L.$ exclusion contours with contours extrapolated from the $68 \% C.L.$ fit (ATLAS only). Colour code per final state: yellow: $\gamma\gamma$, green: $\tau\tau$, red: WW, blue: ZZ. Gray rhombus: SM}
\label{fig:Atl2sig}
\end{center}
\end{figure}

We therefore fitted the parameters of each log-likelihood paraboloid, so that the positions of the points of the contours taken from the plots matched those of the ellipsis of points with fixed log-likelihood equal to $2.3$ , \textsl{i.e.} the value at which lays the $68\%\ C.L.$ of a 2 degrees of freedom (d.o.f.) $\chi^2$ distribution. The comparison of the LHC results and their fitted ellipes is presented on figure \ref{fig:Fit1sig}. In order to further test the validity of the hypothesis of gaussianity of the log-likelihood functions, we compared the $95\%\ C.L.$ exclusion contours extrapolated from the fitted $\chi^2$ distributions to the experimental contours from the ATLAS experiment. The comparison plot is displayed in figure \ref{fig:Atl2sig} and show reasonable agreement.

\subsection{Constraining models}

After obtaining the shape of the likelihood for the signal-strengths, they can be translated into a single likelihood for the parameters defined previously. However, the proper combination of the information from the different channels can only be done by experimentalists, as systematic uncertainties might be correlated, through theoretical uncertainties in the description of the same decay channel or production mode, even between different experiments, and through experimental systematics, within a same experiment. 
Being aware of this, we nevertheless combine the different channels by multiplying the various likelihood functions. Moreover, as the experimental contours can only be interpreted as likelihood-ratios, the most likely point being the null hypothesis, we combine our different functions as such, giving us the likelihood ratio function $\mathcal{L}$, or alternatively the log-likelihood ratio function $\Delta\chi^2$:

\beq
 \mathcal{L}\left(\kappa_j\right)& = &\frac{\prod_i \mathcal{L}_i\left(\kappa_j\right)}{\prod_i \mathcal{L}_i\left(\hat\kappa_j\right)} \\
 \\
 \text{or\hspace{1cm}} \Delta\chi^2\left(\kappa_j\right)&=& \sum_i \chi^2_i\left(\kappa_j\right) - \sum_i\chi_i^2\left(\hat\kappa_j\right)
\eeq

where $\kappa_j$ represents the set of all $\kappa$'s, and $\hat\kappa_j$ their values maximising the total likelihood (minimising the total log-likelihood).

According to Wilks's theorem, the log-likelihood function follows a $\chi^2$ distribution with a number of d.o.f. equal to the number of parameters on which the function depends. In our case, with 6 parameters, the $\Delta\chi^2$ function should therefore follow a 6 d.o.f $\chi^2$ distribution. 

\begin{figure}[ht]
\begin{center}
 \includegraphics[width=0.45\textwidth]{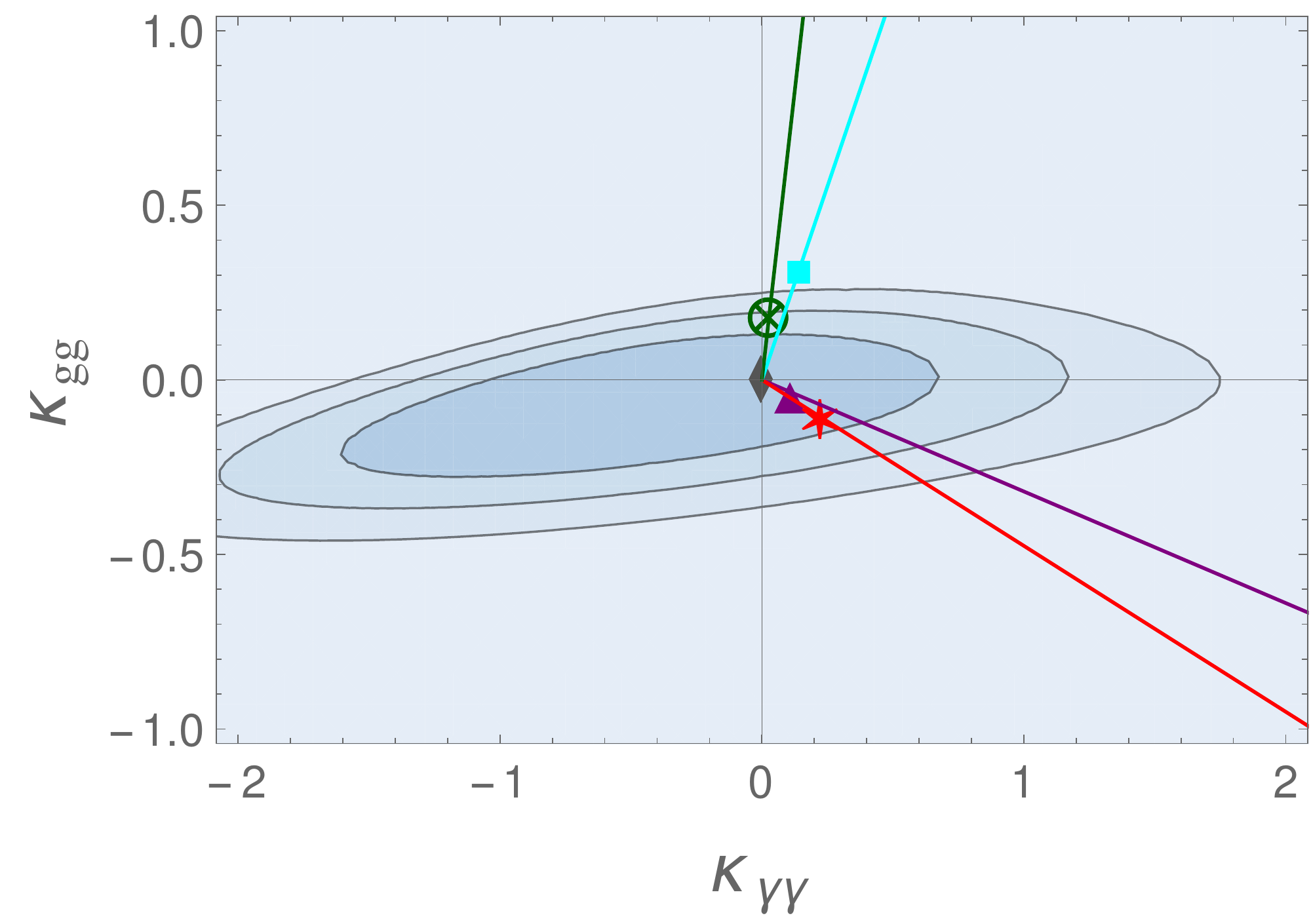} ~~~~~~
 \includegraphics[width=0.45\textwidth]{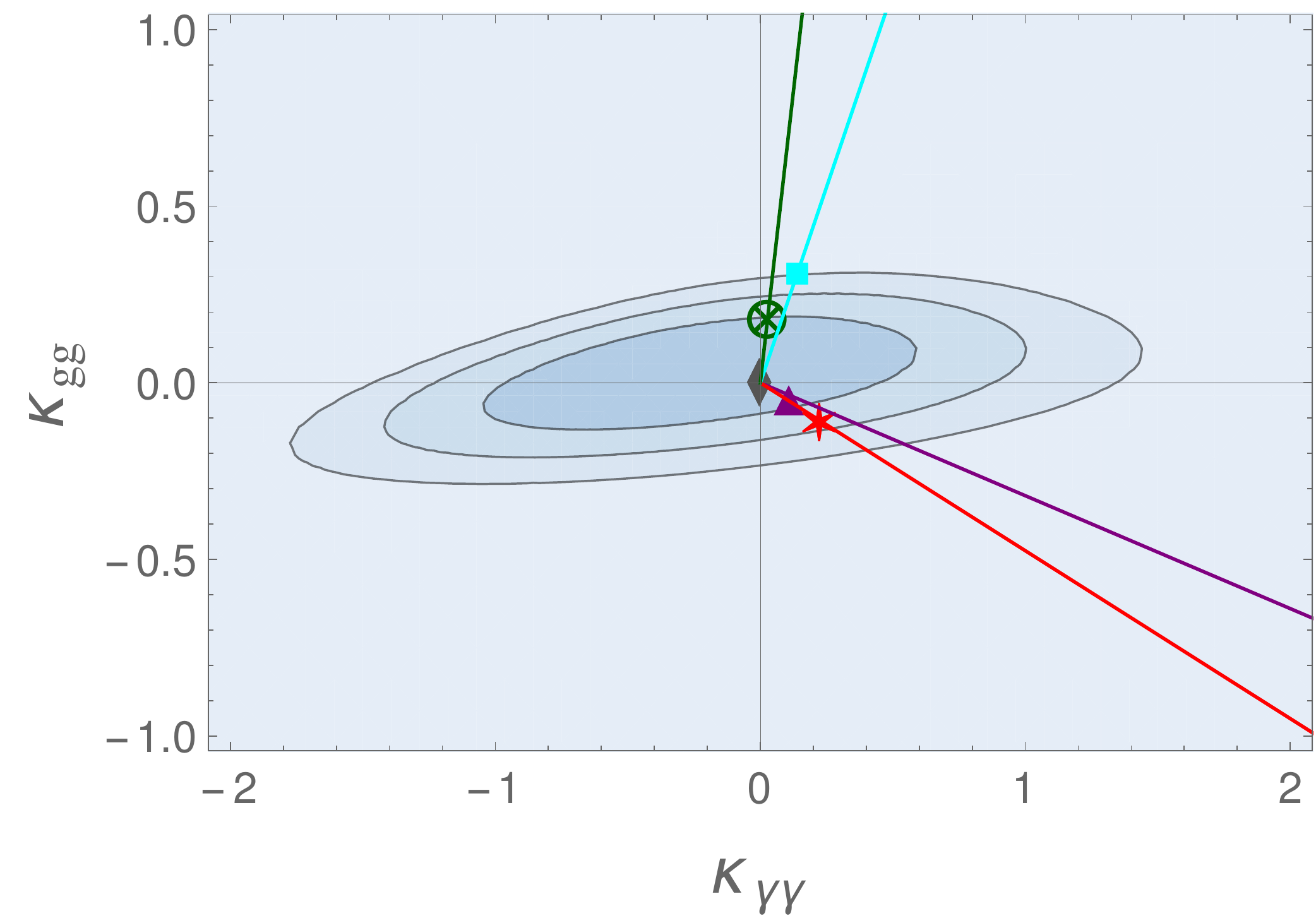}
\caption{Constraints on various models, in the plane spanned by $\kappa_{\gamma\gamma}$ and $\kappa_{gg}$. All other parameters are taken equal to 1. The $\Delta\chi^2$ distribution is compared to $68\%,\ 95\% \text{ and } 99.7\%\ C.L.$ values for a 6 d.o.f. $\chi^2$ distribution. The left figure uses results from former LHC analyses, and the right one uses results from the legacy analyses.}
\label{fig:Const6dof}
\end{center}
\end{figure}

In order to see how this applies to specific models, we provide a few examples of models that can be described by our parametrisation, and for which values of the parameters have been described in \cite{cacciapaglia_0901}:

\begin{itemize}
\item [\textcolor{cyan}{$\blacksquare$}] Colour octet model\cite{PhysRevD.74.035009}: Model where the scalar SM sector is extended with a colour octet. As this octet can be decomposed into SU(2) representations, additional neutral and charged scalars appear. The model point represented in Fig. \ref{fig:Const6dof} corresponds to $m_S=750,\ \lambda_1=4,\ \lambda_2=1,\text{ and }\lambda_3=2 \lambda_2$, while the line corresponds to a varying value of $m_S$.

\begin{center}
\parbox{5 cm}{$$\kappa_{\gamma\gamma} = \frac{3}{2}\frac{\lambda_1 v^2}{4 m_{S^+}^2} $$}     \parbox{7cm}{$$\kappa_{gg} \simeq \frac{C(8)}{2} \frac{(2 \lambda_1 + \lambda_2 )v^2}{4 m_S^2}$$}
\end{center}

\item [\textcolor{vertfonce}{$\otimes$}] 5D UED\cite{PhysRevD.64.035002}: A Universal Extra Dimension model with one extra dimension, where only the top and $W$ Kaluza-Klein resonances contribute and the result scales with the size of the extra dimension. The model point here corresponds to $m_{KK} = 500\ GeV$, and the line corresponds to a varying $m_{KK}$.

\begin{center}
\parbox{9cm}{$$\kappa_{\gamma\gamma} =-\frac{63 \pi^2}{16 \times 6}\left(\frac{m_W}{m_{KK}}\right)^2 + \frac{\pi^2}{6}\left(\frac{m_t}{m_{KK}}\right)^2 $$}     \parbox{6cm}{$$\kappa_{gg} =  \frac{\pi^2}{6}\left(\frac{m_t}{m_{KK}}\right)^2 $$}
\end{center}

\item [\textcolor{purple}{$\blacktriangle$}] Simplest Little Higgs\cite{Schmaltz:2004de}: The electroweak $SU(2)$ gauge group is in this model embedded in a larger $SU(3)$, spontaneously broken by triplets through a Higgs-like mechanism. The corrections scale with the $W'$ mass, and the point corresponds to the value $m_{W'}=500\ GeV$ within electroweak precision constraints for a model with $T$-parity.

\begin{center}
\parbox{8 cm}{$$\kappa_{\gamma\gamma} \simeq \left(\frac{47}{12}-\frac{3}{16}\left(7+A_W(\tau_W)\right))\right)\left(\frac{m_W}{m_{W'}}\right)^2  $$}     \parbox{6cm}{$$\kappa_{gg} \simeq -\frac{4}{3}\left(\frac{m_W}{m_{W'}}\right)^2$$}
\end{center}

\item [\textcolor{red}{$\ast$}] Littlest Higgs\cite{ArkaniHamed:2002qy}: The Higgs in this model is one of a set of pseudo-Goldstone bosons in an
$SU(5)/SO(5)$ nonlinear sigma model. The result scales with the symmetry breaking scale $f$, which can also be set to low values for a model with $T$-parity\cite{Low:2004xc} (there is also a mild dependence on the mass of the extra gauge boson contributing to $\kappa_{\gamma\gamma}$, arbitrarilly fixed to be equal to $f$ and on the triplet VEV $x$, that we set to $x=0$);

\begin{center}
\parbox{15cm}{$$\kappa_{\gamma\gamma} \simeq \frac{(195+64x-73x^2)v^2}{128 f^2} + \frac{9}{16}(7+A_W(\tau_W))\left(\frac{m_W^2}{m_{W'}^2}-\frac{(5-x^2)v^2}{8 f^2}\right)$$}     \parbox{8cm}{$$\kappa_{gg} \simeq -\frac{(7-4x+x^2)v^2}{8 f^2}$$}
\end{center}

\end{itemize}

Figure \ref{fig:Const6dof} represents the contours of the regions excluded by two sets of LHC results, at $68\%,\ 95\% \text{ and } 99.7\%\ C.L.$, in the plane spanned by $\kappa_{gg}\text{ and }\kappa_{\gamma \gamma}$. The SM is represented by a grey rhombus, at the centre of the figure. The right plot was produced using legacy data described before, whereas the left one was produced using the same LHC results as in \cite{Cacciapaglia:2012wb} and is displayed here to see the evolution of the constraints. We can see that the latest bounds are more stringent but also shifted so that the centre is closer to the SM. Therefore, some values of the parameters that were excluded at a given confidence level no longer are, as we will see a bit further.

More precisely, from these constraints, we can compute values of the parameters for which the models are excluded, for instance along the lines drawn on the figure.
In table \ref{tab:ModConst}, four values are provided, corresponding to two different approaches of the way the model should match the observations, each one both for former constraints and the recent legacy ones. 

The first and third lines of the table show the values of parameters excluded when comparing the model's likelihood $\Delta\chi^2$ to a 6-dof $\chi^2$ distribution. This means that we consider that the model in itself does not describe all new physics, and other modifications might have an influence on $\kappa_b,\ \kappa_t,...$, but we slice the parameter space to observe the plane where these parameters are 1. We then get a parameter value under which the model is excluded at $95\%\ C.L.$. 

On the second and fourth lines, we instead consider that the only modification to the Higgs couplings come from the model we are testing. The other $\kappa$'s are then no longer free parameters, and Wilks's theorem asserts that the log-likelihood ratio function now follows a $\chi^2$ distribution with $n$ d.o.f., $n$ being the number of parameters of the model (and the null hypothesis being of course the total $\chi^2$ minimum, amongst points in $\kappa$ space that can be described by model). The bounds are therefore tighter, and the values excluded higher.

\begin{center} 
\begin{table}[ht]\hspace{-1cm}
 \begin{tabular}[ht!]{|c|c|c|c|c|c|}
 \hline & $\chi^2$ & Colour Octet & 5D UED & Simplest Little Higgs & Littlest Higgs\\
 \hline \multirow{2}{1.6cm}{Previous \hspace{0.2cm}analyses} & $6\ d.o.f$ & $m_S = 1040\ GeV$ & $m_{KK} = 480\ GeV$ & $m_{W'} = 200\ GeV$ & $m_{W'} = 475\  GeV$\\ 
 \cline{2-6}  & $n\ d.o.f$ & $m_S = 1220\ GeV$ & $m_{KK} = 660\ GeV$ & $m_{W'} = 280\ GeV$ & $m_{W'} = 540\ GeV$\\\hline
 \hline \multirow{2}{1.6cm}{Legacy \hspace{0.2cm}analyses} & $6\ d.o.f$ & $m_S = 835\ GeV$ & $m_{KK} = 430\ GeV$ & $m_{W'} = 250\ GeV$ & $m_{W'} = 580\  GeV$\\ 
 \cline{2-6} & $n\ d.o.f$ & $m_S = 925\  GeV$ & $m_{KK} = 530\ GeV$ & $m_{W'} = 380\ GeV$ & $m_{W'} = 690\ GeV$\\\hline
  \end{tabular}
\caption{Values of the parameters excluded for different models under specific assumptions. The first two lines correspond to old constraints and the two last lines to legacy constraints. In each case, in the first lines model predictions are interpreted as following a 6-d.o.f $\chi^2$ distribution and in the second line as a $\chi^2$ distribution with number of parameters equal to the number of parameters of the model}
\label{tab:ModConst}
\end{table}
\end{center}

From the values in this table, we can note that the most stringent bounds for given models are not necessarily the most recent ones. For instance, the colour octet and the UED models are less constrained by legacy results than by the former constraints. This can be easily understood by comparing the two diagrams in figure \ref{fig:Const6dof}. The constraints on Little Higgs models, on the other hand, follow the expected variation, and are more constrained by more recent results.

\section*{Conclusion}


The parametrisation of the Higgs boson couplings described here, laying between the simple parametrisations with few parameters and the more complete but complex ones, is complementary to other similar parametrisations, and could be used in its full potential if constrained directly by experiments. In the meantime, the procedure presented here is a straightforward way for theoreticians to get constraints on models including new loops in Higgs physics. As an example, we computed bounds on a set of models with constraints extracted from CMS and ATLAS Run 1 Legacy analyses, showing this parametrisation is a simple and effective way to constrain BSM models.
\vspace{0.2cm}

Note: The ATLAS collaboration released during the writing of this paper a conference note containing a figure summarising all Higgs signal strengths constraints. It can be found in Fig. 3 \textsl{in} \cite{ATLAS-CONF-2015-007}, however, as this figure is marked as a preliminary result, and we did not expect the results to be different from the final analysis papers, we did note include it in our study.

\bibliography{Biblio}

\begin{thebibliography}{10}

\bibitem{ATLAS-CONF-2015-007}
{Measurements of the Higgs boson production and decay rates and coupling
  strengths using pp collision data at $\sqrt{s}$ = 7 and 8 TeV in the ATLAS
  experiment}.
\newblock Technical Report ATLAS-CONF-2015-007, CERN, Mar 2015.

\bibitem{PhysRevD.90.112015}
G.~Aad et~al.
\newblock {Measurement of Higgs boson production in the diphoton decay channel
  in $pp$ collisions at center-of-mass energies of 7 and 8~TeV with the ATLAS
  detector}.
\newblock {\em Phys. Rev. D}, 90:112015, Dec 2014.

\bibitem{PhysRevD.91.012006}
G.~Aad et~al.
\newblock {Measurements of Higgs boson production and couplings in the
  four-lepton channel in $pp$ collisions at center-of-mass energies of 7 and
  8~TeV with the ATLAS detector}.
\newblock {\em Phys. Rev. D}, 91:012006, Jan 2015.

\bibitem{Aad:bbATLAS}
G.~Aad et~al.
\newblock {Search for the bb decay of the Standard Model Higgs boson in
  associated (W/Z)H production with the ATLAS detector}.
\newblock {\em Journal of High Energy Physics}, 2015(1), 2015.

\bibitem{Aad:1975394}
Georges Aad et~al.
\newblock {Observation and measurement of Higgs boson decays to $WW^\ast$ with
  the ATLAS detector}.
\newblock Technical Report arXiv:1412.2641. CERN-PH-EP-2014-270, CERN, Dec
  2014.

\bibitem{Aad:1982276}
Georges Aad et~al.
\newblock {Evidence for the Higgs-boson Yukawa coupling to tau leptons with the
  ATLAS detector}.
\newblock Technical Report arXiv:1501.04943. CERN-PH-EP-2014-262, CERN, Jan
  2015.

\bibitem{PhysRevD.64.035002}
Thomas Appelquist, Hsin-Chia Cheng, and Bogdan~A. Dobrescu.
\newblock Bounds on universal extra dimensions.
\newblock {\em Phys. Rev. D}, 64:035002, Jun 2001.

\bibitem{ArkaniHamed:2002qy}
N.~Arkani-Hamed, A.G. Cohen, E.~Katz, and A.E. Nelson.
\newblock {The Littlest Higgs}.
\newblock {\em JHEP}, 0207:034, 2002.

\bibitem{Cacciapaglia:2012wb}
Giacomo Cacciapaglia, Aldo Deandrea, Guillaume~Drieu La~Rochelle, and
  Jean-Baptiste Flament.
\newblock {Higgs couplings beyond the Standard Model}.
\newblock {\em JHEP}, 1303:029, 2013.

\bibitem{cacciapaglia_0901}
Giacomo Cacciapaglia, Aldo Deandrea, and Jeremie Llodra-Perez.
\newblock {Higgs $\to\ \gamma \gamma$ beyond the Standard Model}.
\newblock {\em JHEP}, 0906:054, 2009.

\bibitem{Khachatryan:2014jba}
Vardan Khachatryan et~al.
\newblock {Precise determination of the mass of the Higgs boson and tests of
  compatibility of its couplings with the standard model predictions using
  proton collisions at 7 and 8 TeV}.
\newblock 2014.

\bibitem{Low:2004xc}
Ian Low.
\newblock {T parity and the littlest Higgs}.
\newblock {\em JHEP}, 0410:067, 2004.

\bibitem{PhysRevD.74.035009}
Aneesh~V. Manohar and Mark~B. Wise.
\newblock {Flavor changing neutral currents, an extended scalar sector, and the
  Higgs production rate at the CERN Large Hadron Collider}.
\newblock {\em Phys. Rev. D}, 74:035009, Aug 2006.

\bibitem{Schmaltz:2004de}
Martin Schmaltz.
\newblock {The Simplest little Higgs}.
\newblock {\em JHEP}, 0408:056, 2004.

\end{thebibliography}

\end{fmffile}
\end{document}